\documentclass[a4paper,11pt]{article}
\pdfoutput=1 % if your are submitting a pdflatex (i.e. if you have
\usepackage{jinstpub} % for details on the use of the package, please
\usepackage{amsthm}
\usepackage{hyperref}
\usepackage{color}
\usepackage{fixltx2e} %%To use subscript and superscript
\usepackage{tikz}
\usepackage{epstopdf}
\usepackage{subfigure}
\usepackage{txfonts}

\newcommand{\sqdiamond}[1][fill=black]{\tikz [x=1.5ex,y=1.5ex,line width=.1ex, yshift=-0.285ex] \draw  [#1]  (0,.5) -- (.5,1) -- (1,.5) -- (.5,0) -- (0,.5) -- cycle;}%
\newcommand{\MyDiamond}[1][fill=black]{\mathop{\raisebox{-0.275ex}{$\sqdiamond[#1]$}}}

\newcommand{\sqtriangle}[1][fill=black]{\tikz [x=1.2ex,y=1.2ex,line width=.1ex, yshift=-0.285ex] \draw  [#1]  (0,1) -- (1,1) -- (0.5,0) -- cycle;}%
\newcommand{\MyTriangle}[1][fill=black]{\mathop{\raisebox{-0.275ex}{$\sqtriangle[#1]$}}}

\newcommand{\MySquare}[1][fill=white]{\tikz [x=1.2ex,y=1.2ex,line width=.1ex, yshift=-0.285ex] \draw  [#1]   (0,1) -- (1,1) -- (1,0) -- (0,0) -- cycle;}%

\title{\boldmath Pressure effect in the X-ray intrinsic position resolution in noble gases and mixtures}

\author[a,1]{C.D.R. Azevedo,\note{Corresponding author.}}
\author[a]{P.M. Correia,}
\author[b]{D. Gonz\'alez-D\'iaz,}
\author[c]{S. Biagi,}
\author[a]{A.L.M. Silva,}
\author[a]{L.F.N.D. Carramate}
\author[a]{and J.F.C.A. Veloso}

\affiliation[a]{I3N - Physics Department, University of Aveiro\\Campus Universit\'ario de Santiago 3810-193 Aveiro, Portugal}
\affiliation[b]{CERN\\ CH-1211 Geneva 23, Switzerland}
\affiliation[c]{Uluda\u{g} University, Faculty of Arts and Sciences\\ Physics Department, Bursa,Turkey}

\emailAdd{cdazevedo@ua.pt}

\abstract{A study of the gas pressure effect in the position resolution of an
interacting X- or gamma-ray photon in a gas medium is
performed. The intrinsic position
resolution for pure noble gases (Argon and Xenon) and their mixtures
with CO\textsubscript{2} and CH\textsubscript{4} were calculated for several gas pressures (1-10
bar) and for photon energies
between 5.4 and
60.0 keV, being possible to establish a linear match between the
intrinsic position resolution and the inverse of the gas pressure in
that energy range.

In order to evaluate the quality of the method here described, a
comparison between the available experimental data and the calculated
one in this work, is done and discussed. In the majority of the cases,
a strong agreement is observed.}

\keywords{Gaseous detectors; Gaseous imaging and tracking detectors; Charge transport and multiplication in gas; Detector modelling and simulations I (interaction of radiation with matter, interaction of photons with matter, interaction of hadrons with matter, etc)}

\arxivnumber{1605.06256} % only if you have one

\begin{document}
\maketitle
\flushbottom

\section{Introduction}
\label{introduction}

Previous works have shown that the intrinsic position resolution for X
and gamma-ray detection in a gas medium depends mainly on the photon
energy and on the gas choice \cite{ref1}.
Moreover, it shows that the position resolution reaches a minimum value
when the photon energy is slightly higher than the atom K-shell energy
of the gas.

The main drawback when using gas as radiation detection medium arises
from the low photon detection efficiency. The most usual way to
overcome this problem is to increase the gas pressure, thus increasing
the number of gas atom/molecules per unit volume. Also, the pressure
increase will benefit the energy resolution and light production when
electroluminescence measurement is needed \cite{ref2,ref3,ref4}.
It is expected that the pressure increase will reduce the electrons
diffusion in the gas medium, which will in turn result in a better
position resolution when compared to the value obtained at atmospheric
pressure \cite{ref5,ref6}. Nevertheless, secondary
effects such as X-ray fluorescence process, can play a major role on
the position resolution, thus improving or degrade the resolution value
depending on the geometry \cite{ref1}.

Experimental measurements have shown that the increase of gas pressure
leads to an improving of the position
resolution \cite{ref2,ref7,ref8}, however the measurement technique relies on detectors
where the position resolution measured depends on the detector gain and
on the electron diffusion in the gas. Thus, in order to decouple the
position resolution from the detector gain, the intrinsic position
resolution of pressurized pure noble gases
(Ar and Xenon, in the pressures range 1-10\,bar) was calculated, studied and compared
with the available experimental data.

In order to compare the obtained simulated results with the available
experimental data \cite {ref7,ref8,ref9},
mixtures of pure noble gases with molecular VUV-quenchers, namely
CO\textsubscript{2}, and in the case of Ar,
also with CH\textsubscript{4,} were included in this study.

\section{Method}
\label{method}

For the calculations a software tool which includes secondary processes
like X- ray fluorescence, Auger, Coster-Kronig and Shake-off was used:
Degrad \cite{ref10}.
This software tool, developed by S. Biagi, is at the moment able to
calculate the atomic cascade initiated by X-ray photons or electrons
interacting in the gas, returning the number of ionizations,
excitations and the position of each thermalized electron \cite{ref11}.

For each simulated interaction, the detection position was defined as
the center-of-gravity of the generated charge. In the absence of a
detailed topological analysis of the ionization trails (as, e.g. in \cite{ref12},
such a definition likely represents the ultimate accuracy limit in most
practical X-ray gas detectors.

An image of the obtained centroid positions of each individual charge
cluster was constructed and an analysis chain similar to the method
present in \cite{ref1} was applied to the results. The position resolution was taken as the
Full Width at Half Maximum (FWHM) of a Gaussian function fitted to the
projected data of the constructed image, following the Line Spread Function (LSF) method used in \cite{ref13}.

For each condition, 500.000 events were generated using Degrad2.13. All
the events were considered to interact in (X,Y,Z)=(0,0,0) in order to
simulate an infinitesimal point-like interaction. An electric field of
300\,V.cm\textsuperscript{-1}.bar\textsuperscript{-1}
was considered. The gas temperature was set to 20\,{\textordmasculine}C
and, in order to speed up the calculations, the electrons were
considered thermalized when their energy follows 1 eV bellow the lowest
excitation energy of the main gas in the mixture. Compton events are
negligible for the studied energy range, therefore they were not
considered in the analysis.

\section{Results and discussion}
\label{Results}

The position resolution in pure Noble gases and their mixtures with
CH\textsubscript{4} and CO\textsubscript{2}
(Ar/10\%CH\textsubscript{4}, Ar/20\%CO\textsubscript{2}, Ar/30\%CO\textsubscript{2},
Xe/10\%CO\textsubscript{2}, pure Argon and
pure Xenon) in pressures ranging from 1 to 10\,bar for photon
interactions with energies between 5.4 and 60.0\,keV were calculated
considering an infinite gas volume.
\begin{figure}[htbp]
\centering \includegraphics[width=0.6\textwidth]{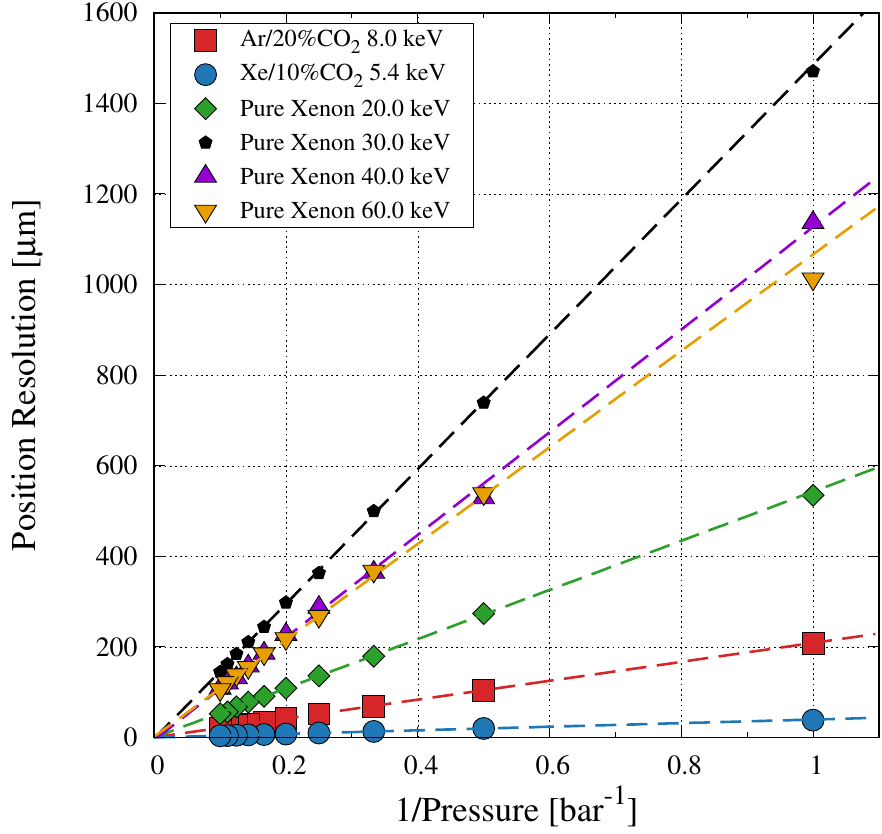}
\caption{Position resolution as a function of the inverse of the pressure for different photons in Argon and Xenon mixtures with CO\textsubscript{2}, considering an infinite volume. Dashed lines are the fits of a linear function to the calculated data points.}
\label{Figure1}
\end{figure}
Figure~\ref{Figure1} presents the obtained results(where error
bars are smaller than the data point size) plotted as function of the
inverse of the pressure. By fitting the data
with a linear function (dashed lines) it becomes clear the linear
relation between
the calculated position resolution and the inverse of the gas
pressure, as already referred in \cite{ref7}.

\begin{equation}
FWHM\propto \frac{1}{P}
\label{eq1}
\end{equation}

This demonstrates that it becomes possible to calculate the position
resolution for a given pressure by just dividing the position
resolution value at 1 bar (see results in \cite{ref1})
by the pressure value.

\begin{figure}[htbp]
\centering 
\subfigure[]{\label{Figure2}\includegraphics[width=0.45\textwidth]{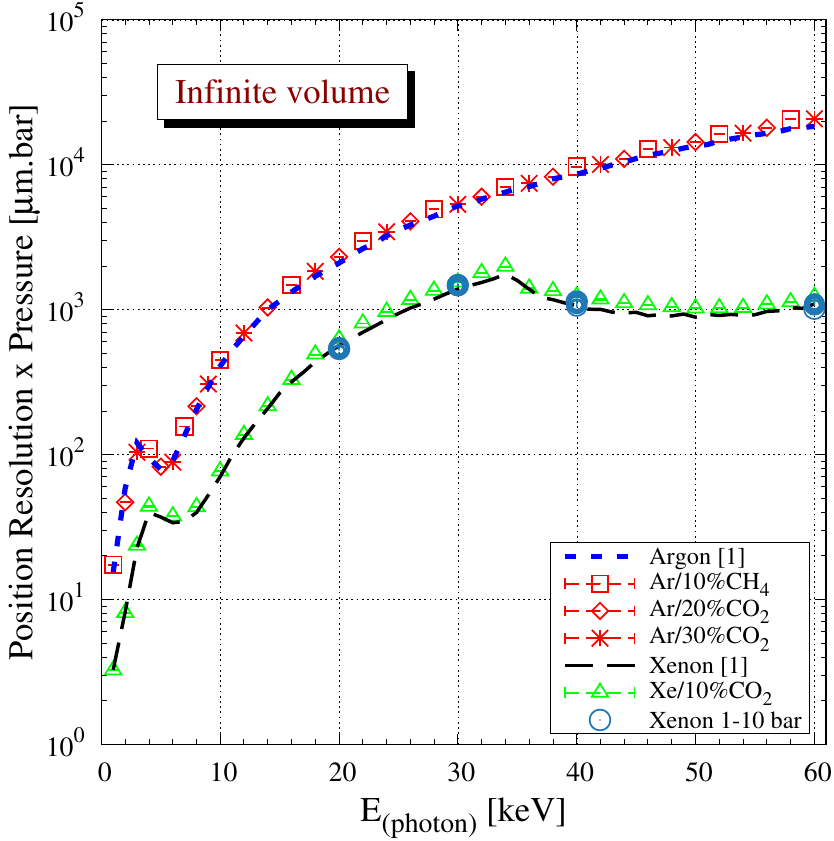}}
\subfigure[]{\label{Figure3}\includegraphics[width=0.45\textwidth]{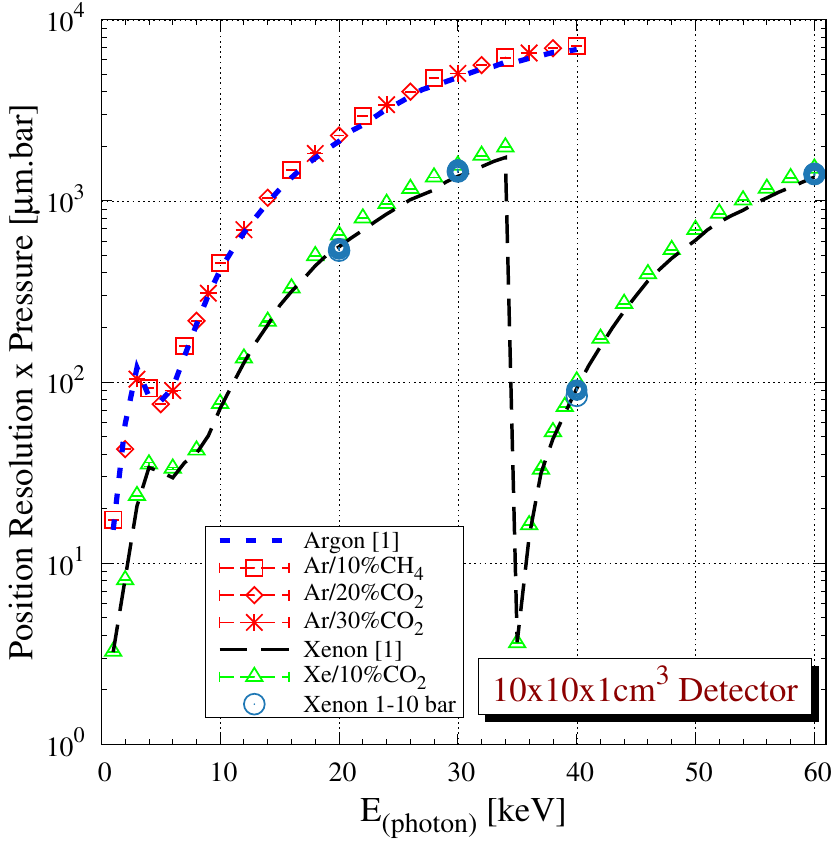}}
\caption{Position resolution multiplied by pressure value as a function of the photon energy for: a) an infinite volume; b) 10x10x1\,cm\textsuperscript{3} detector. All data points, except (open circles \textcolor[rgb]{0.0,0.5176471,0.81960785}{O}), were obtained at 1 bar.}
\label{Figure2Tot}
\end{figure}

Another exercise is to multiply the position resolution by the pressure
value: the calculated data points will be overlaid for all the
pressures. That assumption is verified in Figure~\ref{Figure2}
and plotted as a function of the photon energy. In this plot we can
also observe a small degradation of the position resolution depending
on the fraction of quencher in the gas mixture.

The explanation for the data curves profile shown in
Figure~\ref{Figure2Tot} are
explained in ref \cite{ref1}
and are related to the \textit{K} and \textit{L} shells of the used gas.

In order to infer if this scaling still valid for a finite gas volume,
calculations were performed on a typical volume of
10x10x1\,cm\textsuperscript{3} volume. The results are presented in
Figure~\ref{Figure3} where the $1/P$ scaling can be again observed
by multiplying the position resolution by the pressure value. A minor
degradation of the position resolution value dependence with the
fraction of quencher in the gas mixture is again observed for the used
energy range.

\section{Comparison with experimental data}
\label{comparsion}

\begin{figure}[htbp]
\centering 
\subfigure[]{\label{Figure4}\includegraphics[width=0.45\textwidth]{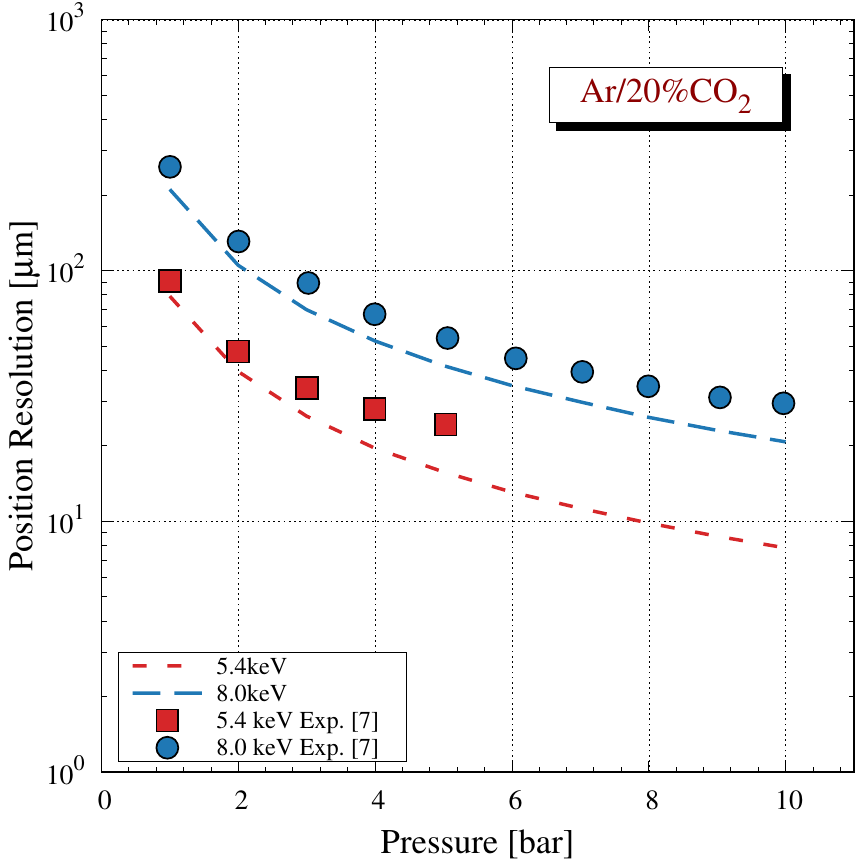}}
\subfigure[]{\label{Figure5}\includegraphics[width=0.45\textwidth]{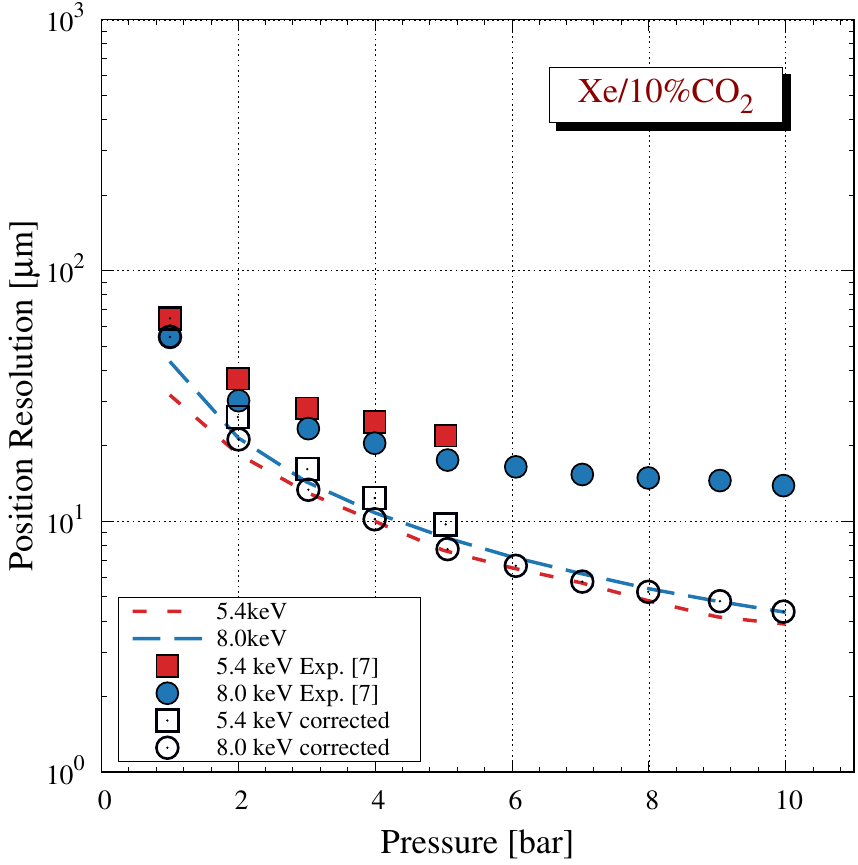}}
\caption{Position Resolution multiplied by the pressure as a function of photon
energy: a) Argon/20\%CO\textsubscript{2}; b) Xe/10\%CO\textsubscript{2}. Lines are calculations at 1\,bar (this work) while points are experimental results from \cite{ref7,ref8,ref9}}
\label{Figure3Tot}
\end{figure}

The calculated results for a 10x10x1\,cm\textsuperscript{3} volume were
compared with the available experimental data for
Xe/10\%CO\textsubscript{2} and
Ar/20\%CO\textsubscript{2} from ref \cite{ref7}
and are shown in Figure~\ref{Figure3Tot}.

It can be observed a good agreement between the experimental and the
calculated data for
Ar/20\%CO\textsubscript{2} mixtures. The
slightly better values achieved in the simulation can be explained
since we are just calculating the intrinsic gas position resolution,
i.e., other parameters like the electron diffusion in gas along the
drift distance were not taken into account.

For Xe/10\%CO\textsubscript{2} the
experimental and calculated data are not in as good agreement compared
to the previous mixture, mainly due to the experimental data deviation
from the $1/P$ behaviour. In fact the experimental data follows the
inverse of the pressure square-root showing that the position
resolution for such small values is being dominated by the electrons
diffusion in the gas \cite{ref5}
(that was not taken into account for the calculations). This is also
one of the reasons for the deviations pointed in \cite{ref7}
where the authors expect a value of 6\,$\mu$m for the position
resolution, which is in accordance to our calculations. In
Figure~\ref{Figure5} is also shown the experimental values
corrected to the diffusion effect (closed data points) by multiplying
the data by the square root of the pressure. In this case we can
observe a good agreement between the experimental corrected data and
this work for both, 8.0\,keV and 5.4\,keV photons. We believe that this
fact could be again being related to experimental conditions.

\begin{figure}[htbp]
\centering 
\subfigure[]{\label{Figure6}\includegraphics[width=0.45\textwidth]{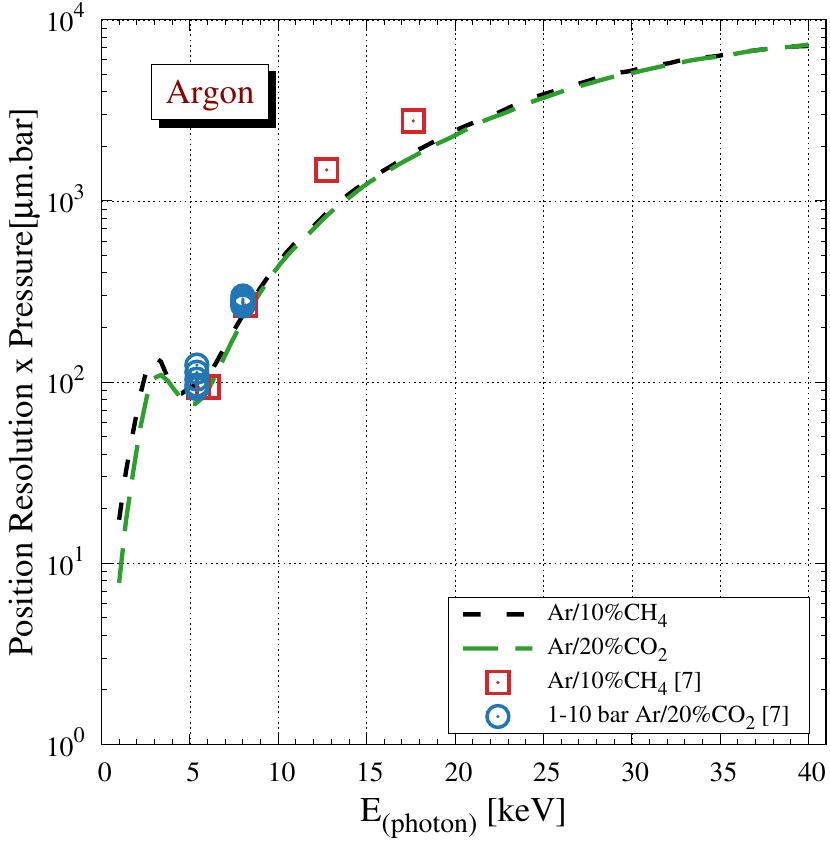}}
\subfigure[]{\label{Figure7}\includegraphics[width=0.45\textwidth]{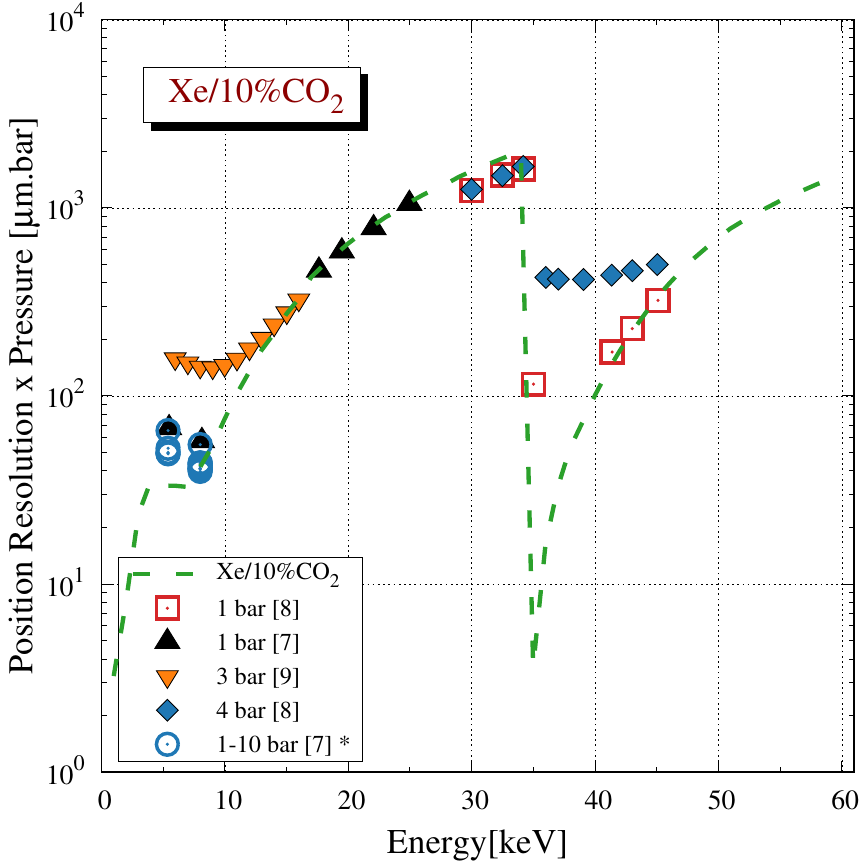}}
\caption{Position Resolution multiplied by the pressure as a function of photon
energy for: a) typical Argon mixtures; b) Xe/10\%CO\textsubscript{2}. Lines are calculations at 1\,bar
(this work) while points are experimental results from \cite{ref7,ref8,ref9}.
*The experimental values in b) were deconvoluted from the electron diffusion prior to the multiplication by the pressure.}
\label{Figure4Tot}
\end{figure}

The experimental position resolution data from Figure~\ref{Figure3} and 
Figure~\ref{Figure5} (corrected values) plus the data obtained in refs
\cite{ref8,ref9}, were multiplied by the pressure and are presented in
Figure~\ref{Figure4Tot}. Again, a
good agreement between the calculated and the experimental data is
observed, with exception for low energy photons. This can be explained
by the experimental limitations: low energetic photons present low
signal-to-noise ratios and are affected by the electronic noise, thus
degrading the position resolution as explained in \cite{ref7}.
As the photon energy increases, the signal-to-noise ratio also
increases, improving the agreement (Figure~\ref{Figure7}, 3\,bar
experimental data set - $\MyTriangle[draw=black,fill=orange]$).
Another difference between both data is observed after the Xenon
\textit{K}-shell (34.6\,keV). For the 1\,bar data set the explanation is the same as pointed
before. After the \textit{K}-shell, the photoelectron kinetic energy will
decrease leading to the emission of a fluorescence photon that has high
escaping probability for relatively slim detectors, thus degrading the
position resolution. This means that, in such case, the detector will
only detect the small energy of the photoelectron being again affected
by the electronic noise. When the photon energy further increases the
position resolution will approach the intrinsic: 1 and 4\,bar data sets
($\MySquare[draw=red,ultra thick]$,$\MyDiamond[draw=black,fill=cyan]$) in Figure~\ref{Figure7}.
The values for the experimental data sets 1-10\,bar (
\textcolor{blue}{O}) in Figure~\ref{Figure7} were deconvoluted
from the electron diffusion prior to the multiplication by the pressure
(see Figure~\ref{Figure5} discussion).

\section{Conclusions}
\label{Conclusions}

This work presents the influence of the gas pressure on the intrinsic
position resolution for pure Ar, Xe and mixtures with
CO\textsubscript{2} and
CH\textsubscript{4}. A linear behavior
between the position resolution and the inverse of pressure was
observed for pressures between 1-10 bar and for photon energies ranging
from 5.4-60.0\,keV.

A minor position resolution degradation when using mixture with
CO\textsubscript{2} percentages as high as
30\% was also observed.

A good agreement between the calculated and experimental data was
observed. In the case of Xe/10\%CO\textsubscript{2} mixtures,
deviations from the experimental data were observed, explained by the
electron diffusion in the experimental data, that was not taken into
account during the calculations.

For low energy photons, the deviations between experimental and
calculated data are explained by the influence of the signal-to-noise
ration and its influence on the degradation of the experimental
position resolution.

\acknowledgments

This work was supported by project PTDC/FIS-NUCL/2525/2014 through
FEDER, COMPETE and FCT (Lisbon) programs. C.D.R. Azevedo and A.L.M.
Silva were supported by PostDoctoral grants from FCT (Lisbon)
SFRH/BPD/79163/2011 and SFRH/BPD/109744/2015,
respectively. P.M.M. Correia and L.F.N.D. Carramate were supported by
the FCT (Lisbon) scholarships BD/52330/2013 and SFRH/BD/71429/2010,
respectively.

The research was done within the CERN RD51 Collaboration.

% We suggest to always provide author, title and journal data:
% in short all the informations that clearly identify a document.


\begin{thebibliography}{99}


\bibitem{ref1} C.D.R. Azevedo et al., \emph{Position resolution limits in pure noble gaseous detectors for X-ray energies from 1 to 60 keV}, PLB 741 (2015) 272.
\bibitem{ref2} C.D.R. Azevedo et al., \emph{2D-sensitive hpxe gas proportional scintillation counter concept for nuclear medical imaging purposes}, JINST 6 (2011) C01067.
\bibitem{ref3} C.D.R. Azevedo et al., \emph{A Gaseous Compton Camera using a 2D-sensitive gaseous photomultiplier for Nuclear Medical Imaging}, NIM A 732 (2013) 3.
\bibitem{ref4} J. Renner et al., \emph{Ionization and scintillation of nuclear recoils in gaseous xenon}, NIM A 793 (2015) 62.
\bibitem{ref5} D. Gonz\'alez-D\'iaz et al., \emph{Accurate $\gamma$ and MeV-electron track reconstruction with an ultra-low diffusion Xenon/TMA TPC at 10atm}, NIM A 804 (2015) 8.
\bibitem{ref6} P. Ferrario et al., \emph{First proof of topological signature in the high pressure xenon gas TPC with electroluminescence amplification for the NEXT experiment}, JHEP 2016 (2016) 104.
\bibitem{ref7} J. Fischer et al., \emph{X-ray position detection in the region of 6 $\mu$m RMS with wire proportional chambers}, NIM A 252 (1986) 239.
\bibitem{ref8} G.C. Smith et al., \emph{X-ray position resolution in proportional chambers in the region of 100 $\mu$m (FWHM) above the xenon K-edge}, NIM A 350 (1994) 621.
\bibitem{ref9} T.J. Shin et al., \emph{Two-dimensional multiwire gas proportional detector for X-ray photon correlation spectroscopy of condensed matter}, NIM A 587 (2008) 434.
\bibitem{ref10} S. Biagi, Degrad. \url{http://consult.cern.ch/writeup/magboltz/}, Last Access: 17-05-2016
\bibitem{ref11} S. Biagi, \emph{Program DEGRAD an accurate auger cascade model for interaction of photons and particles with gas mixtures in electric and magnetic fields (2013)}. \url{http://indico.cern.ch/event/245535/contributions/531797/attachments/420757/584265/cern.rd51.2013.pdf}, Last Access: 17-05-2016
\bibitem{ref12} D. Pfeiffer et al., \emph{First measurements with new high-resolution gadolinium-GEM neutron detectors}, JINST 11 (2016) P05011.
\bibitem{ref13} S.W. Smith, \emph{Special Imaging Techniques}, in The Scientist and Engineer's Guide to Digital Signal Processing, Second., Elsevier Science, 1999, 423.


% Please avoid comments such as "For a review'', "For some examples",
% "and references therein" or move them in the text. In general,
% please leave only references in the bibliography and move all
% accessory text in footnotes.

% Also, please have only one work for each \bibitem.


\end{thebibliography}
\end{document}